\documentclass[amsmath,amssymb,prc,final,showpacs,twocolumn]{revtex4-1}
 
\usepackage{bm,hyphenat,xspace}
\usepackage{graphicx,epsfig}

\newcommand {\mbf}[1]{{\mathbf{#1}}}

\newcommand {\mcu}{\mathcal{U}}

\newcommand{\He}{{}^3\mathrm{He}}
\newcommand{\Hh}{{}^3\mathrm{H}}
\newcommand{\nH}{n\text{-}{}^3\mathrm{H}}
\newcommand{\pHe}{p\text{-}{}^3\mathrm{He}}

\begin{document}

\title {${}^3$H production via neutron-neutron-deuteron recombination}
 
\author{A.~Deltuva} 
\affiliation{Centro de F\'{\i}sica Nuclear da Universidade de Lisboa, 
P-1649-003 Lisboa, Portugal }

\author{A.~C.~Fonseca} 
\affiliation{Centro de F\'{\i}sica Nuclear da Universidade de Lisboa, 
P-1649-003 Lisboa, Portugal }

\received{\today}

\begin{abstract}
We study the recombination of two neutrons and deuteron into 
neutron and ${}^3$H using realistic nucleon-nucleon potential models.
Exact  Alt, Grassberger, and Sandhas equations for the 
four-nucleon transition operators are solved in the momentum-space
framework using the complex-energy method with special integration
weights. We find that at astrophysical or laboratory
neutron densities the production of  ${}^3$H via 
the  neutron-neutron-deuteron recombination is much slower 
as compared to the radiative neutron-deuteron capture.
We also calculate neutron-${}^3$H  elastic and total cross sections.
\end{abstract}

 \maketitle

\section{Introduction \label{sec:intro}}

The nonrelativistic quantum mechanics solution of the four-nucleon scattering 
problem has, in the past five years, reached a level of
sophistication and numerical accuracy that makes it a natural theoretical 
laboratory to study nucleon-nucleon (NN) force models
with the same confidence as one has used the three-nucleon system in the past. 
This has been demonstrated in a recent benchmark for $\nH$
and $\pHe$ elastic scattering observables \cite{viviani:11a}, 
where three different theoretical frameworks have been compared, 
namely, the hyperspherical harmonics (HH) expansion method
\cite{viviani:01a,kievsky:08a}, the Faddeev-Yakubovsky (FY) equations 
\cite{yakubovsky:67}  for the wave function components in coordinate space
\cite{lazauskas:04a,lazauskas:09a}, and
the Alt, Grassberger and Sandhas (AGS) equations 
\cite{grassberger:67} for transition matrices 
that were solved in the momentum space \cite{deltuva:07a,deltuva:07b}.
 All methods include not only the hadronic NN interaction,
but also the Coulomb repulsion between protons. 
While the first two methods have the advantage of being able to deal with
charged-particle reactions at very low energies 
and include static three-nucleon forces (3NF), the third one is the only 
method so far to make predictions for multichannel reactions such as 
$d+d \to d+d$, $d+d \to n+\He$, $d+d \to p+\Hh$, and
$p+\Hh \to n+\He$ (and the corresponding inverse reactions)
\cite{deltuva:07c,deltuva:10a}.

In a previous publication \cite{deltuva:12c}
 a major step was taken in extending the AGS 
calculations above three- and four-cluster breakup thresholds. Owing to the
complicated analytic structure of the four-body kernel above breakup 
threshold the calculations were performed using the complex energy method
 \cite{kamada:03a,uzu:03a} whose accuracy and practical applicability
was greatly improved by a special integration method
 \cite{deltuva:12c}. This allowed us  to achieve fully
converged results for  $\nH$ elastic scattering
with realistic NN interactions. We note that the FY calculations
of $\nH$ elastic scattering have been recently extended as well
to energies above the four-nucleon breakup threshold
\cite{lazauskas:12a}, however, using a semi-realistic
NN potential  limited to $S$-waves.

In the present work we extend the method of  Ref.~\cite{deltuva:12c}
to calculate the neutron-neutron-deuteron $(nnd)$ recombination into 
$n+\Hh$ and its time-reverse reaction,
i.e., the three-cluster breakup $n+\Hh \to n+n+d$. 
Although breakup reactions are usually measured in nuclear physics,
the recombination has the advantage that its rate is finite
at threshold where the breakup cross section vanishes due to
phase-space factors. Furthermore,
$n+n+d \to n+\Hh$ is the only hadronic recombination reaction
in the four-nucleon system that at threshold
is not suppressed by the Coulomb barrier (like 
 $n+p+d \to p+\Hh$) or Pauli repulsion (like $n+n+n+p \to n+\Hh$).
It can take place in any environment with neutrons and deuterons
and, with respect to the tritium synthesis, it may be competitive to the 
electromagnetic capture reaction $n+d \to \gamma+\Hh$.
Thus, one may rise the question at what conditions 
the $n+n+d \to n+\Hh$ recombination would dominate over
the  $n+d \to \gamma+\Hh$ radiative capture
and to what extent it is relevant for astrophysical processes.

In addition, we also present results for the $n+\Hh$ elastic scattering
and study the energy dependence of the total  $n+\Hh$ cross section.

\section{4N scattering Equations \label{sec:eq}}

We use the time-reversal symmetry to relate the $nnd$ recombination amplitude
to the three-cluster breakup amplitude of the initial $\nH$ state, i.e.,
\begin{equation} \label{eq:trev}
\langle \Phi_{1} |  T_{13} | \Phi_{3} \rangle =
\langle \Phi_{3} |  T_{31} | \Phi_{1} \rangle.
\end{equation}
Here $| \Phi_{1} \rangle$ is the $\nH$ channel state and 
$| \Phi_{3} \rangle$ is the $nnd$ channel state. The advantage is that
the three-cluster breakup amplitude 
$\langle \Phi_{3} |  T_{3 1} | \Phi_{1} \rangle$
is more directly related to the AGS transition operators $\mcu_{\beta \alpha}$
 calculated in our previous works \cite{deltuva:07a,deltuva:12c}.
Since we use the isospin formalism where 
the nucleons are treated as identical fermions, there are only two 
 distinct  two-cluster partitions, namely,
 $\beta,\alpha=1$ corresponds to the $3+1$ partition (12,3)4
whereas  $\beta=2$  corresponds to the $2+2$ partition (12)(34).
For the initial $\nH$ state we need only $\mcu_{\beta 1}$, i.e., 
we solve the AGS equations for the four-nucleon transition operators 
\begin{subequations}  \label{eq:AGS}   
\begin{align}  
\mcu_{11}  = {}&  -(G_0 \, t \, G_0)^{-1}  P_{34} -
P_{34} U_1 G_0 \, t \, G_0 \, \mcu_{11}  \nonumber \\ 
{}& + U_2   G_0 \, t \, G_0 \, \mcu_{21}, \label{eq:U11}  \\
\label{eq:U21}
\mcu_{21} = {}&  (G_0 \, t \, G_0)^{-1}  (1 - P_{34})
+ (1 - P_{34}) U_1 G_0 \, t \, G_0 \, \mcu_{11}.
\end{align}
\end{subequations}
The free resolvent with the complex energy parameter 
$Z = E+ i\varepsilon$ and the free Hamiltonian $H_0$ is 
\begin{gather}\label{eq:G0}
G_0 = (Z - H_0)^{-1}
\end{gather} 
whereas the pair (12) transition matrix
\begin{gather} \label{eq:t}
t = v + v G_0 t
\end{gather} 
 is  derived from the respective potential $v$.
The  3+1 and 2+2 subsystem transition operators are obtained
from the integral equations
\begin{gather} \label{eq:AGSsub}
U_\alpha =  P_\alpha G_0^{-1} + P_\alpha t\, G_0 \, U_\alpha.
\end{gather}
The basis states are antisymmetric under exchange of the two nucleons 
(12). In the $2+2$ partition the basis states  have to be antisymmetric
also  under exchange of the two nucleons (34).
The full antisymmetry as required for the four-nucleon system is ensured by the 
permutation operators $P_{ab}$ of nucleons $a$ and $b$ with
$P_1 =  P_{12}\, P_{23} + P_{13}\, P_{23}$ and $P_2 =  P_{13}\, P_{24}$.

The $\nH$ elastic and inelastic reaction amplitudes at the available  
energy $E = \epsilon_1 + p_1^2/2\mu_1$  are obtained
in the limit $\varepsilon \to +0$. 
Here  $\epsilon_1$ is the $\Hh$ ground state energy,
$\mbf{p}_1$ is the relative $\nH$ momentum, and
$\mu_1 = 3m_N/4$, $m_N$ being the nucleon mass.
The elastic scattering amplitude is calculated in 
Refs.~\cite{deltuva:07a,deltuva:12c}. The amplitude
for the $nnd$  breakup is obtained by the
antisymmetrization of the general three-cluster breakup amplitude
\cite{deltuva:12e}, resulting
\begin{equation} \label{eq:U0}
\begin{split}  
 \langle \Phi_{3} |  T_{3 1} | \Phi_{1} \rangle 
= {}&  \sqrt{3}  \langle \Phi_{3} | 
[(1- P_{34}) U_1 G_0 \, t \, G_0 \, \mcu_{11} \\ & {} +  
U_2 G_0 \,  t \, G_0 \, \mcu_{21} ]
| \phi_{1} \rangle .
\end{split}
\end{equation}
Here $| \phi_{1} \rangle$ is the Faddeev component of the $\nH$
channel state $| \Phi_{1} \rangle = (1+P_1)| \phi_{1} \rangle$;
$\epsilon_1$ and $| \phi_{1} \rangle$ are obtained by solving the 
bound-state Faddeev equation
\begin{gather}\label{eq:phi}
|\phi_{1} \rangle = G_0 t P_1 |\phi_{1} \rangle
\end{gather} 
at $\varepsilon \to +0$.

We solve the AGS equations \eqref{eq:AGS} in the momentum-space
partial-wave framework. The momentum and angular momentum part of the 
basis states are
$ | k_x \, k_y \, k_z   
[l_z (\{l_y [(l_x S_x)j_x \, s_y]S_y \} J_y s_z ) S_z] \,\mathcal{JM} \rangle$ 
for the $3+1$ configuration and 
$|k_x \, k_y \, k_z  (l_z  \{ (l_x S_x)j_x\, [l_y (s_y s_z)S_y] j_y \} S_z)
\mathcal{ J M} \rangle $ for the $2+2$.
Here  $k_x , \, k_y$, and $k_z$ are the four-particle Jacobi momenta
as given in Ref.~\cite{deltuva:12a}, 
$l_x$, $l_y$, and $l_z$ are the corresponding orbital angular momenta,
$j_x$ and $j_y$ are the total angular momenta of pairs (12) and (34),
$J_y$ is the total angular momentum of the (123) subsystem,
 $s_y$ and $s_z$ are the spins of nucleons 3 and 4, 
 $S_x$, $S_y$, and $S_z$ are the channel spins
of two-, three-, and four-particle systems,
and $\mathcal{J}$ is the  total angular momentum 
with the projection  $\mathcal{M}$.
We include a large number of four-nucleon partial 
waves,  $l_x,l_y,l_z,j_x,j_y,J_y \le 4$ and $\mathcal{J} \le 5$,
such that the results are well converged.
The complex-energy method \cite{kamada:03a}
with special integration weights \cite{deltuva:12c} is used 
to treat the singularities of the AGS equations \eqref{eq:AGS}.
To obtain accurate results for the breakup amplitude
$\langle \Phi_{3} |  T_{3 1} | \Phi_{1} \rangle$ near
the $nnd$ threshold we had to use $0.1$ MeV $ \le \varepsilon \le 0.4$ MeV
that are smaller than  $1.0$ MeV $ \le \varepsilon \le 2.0$ MeV
used in the 
 elastic scattering calculations of Ref.~\cite{deltuva:12c}.
However, the need for relatively small $\varepsilon$ values caused
 no technical problems since the integration with special 
weights \cite{deltuva:12c} provides very accurate treatment
of the $\Hh$ pole whereas 
the quasi-singularities due to deuteron pole are located in a very narrow
region with very small weight, such that about 30 grid points for the 
discretization of each momentum variable were sufficient.

\section{Results \label{sec:res}}

The $nnd$ recombination rate $K_3$ is defined such that the number of 
recombination events per volume and time is $K_3  \rho_n^2 \rho_d$
with $\rho_n$ ($\rho_d$) being the density of neutrons (deuterons).
We calculate it as a function of the relative $nnd$ kinetic energy 
$E_3 = E - \epsilon_d$, i.e.,
\begin{equation} \label{eq:K3}
\begin{split}  
K_3 = & {} \frac{ (2\pi)^7 \mu_1 p_1}
{g_3 \pi^2 (\mu_{\alpha y}\mu_{\alpha})^{3/2} E_3^2} 
 \sum_{m_s} \int d^3k_y \, d^3k_z \\ & {} \times
| \langle \Phi_{3} |  T_{3 1} | \Phi_{1} \rangle |^2 \, 
\delta \left(E_3 - \frac{k_y^2}{2\mu_{\alpha y}} - \frac{k_z^2}{2\mu_{\alpha}} 
\right).
\end{split}
\end{equation}
 Here $\epsilon_d = -2.2246$ MeV is the deuteron bound state energy,
$\mu_{\alpha y}$ and $\mu_{\alpha}$ are the reduced masses associated
with the four-nucleon Jacobi momenta  $k_y$ and $k_z$.
For example, in the 2+2 configuration $k_y$ is the relative
momentum of the two neutrons while $k_z$ is the relative momentum
between the center of mass (c.m.) of the two-neutron subsystem and the deuteron.
The $nnd$ state can be represented in both 3+1 and 2+2 configurations equally 
well;  $\mu_{\alpha y}\mu_{\alpha} = m_N^2/2$. 
The sum in Eq.~\eqref{eq:K3} runs over all initial and final spin projections
$m_s$ that are not explicitly indicated in our notation for the channel states
while $g_3=12$ takes care of the spin averaging in the initial $nnd$ state.
The integral in  Eq.~\eqref{eq:K3}, up to a factor, 
determines also the total cross section $\sigma_3$ for the
three-cluster breakup of the initial $\nH$ state. Thus, the 
$nnd$ recombination rate can be related to  $\sigma_3$ as
\begin{equation} \label{eq:K3s3}
K_3 = \frac{8 \pi g_1 p_1^2}
{g_3 (\mu_{\alpha y}\mu_{\alpha})^{3/2}\,E_3^2 } \, \sigma_3
\end{equation}
where $g_1=4$ is the number of $\nH$ spin states. 
Below the four-nucleon breakup threshold  $\sigma_3$ can be obtained
via the optical theorem as a difference between the total and elastic
cross sections.
The equation \eqref{eq:K3s3}
cannot be used right at the $nnd$
threshold where both $E_3$ and  $\sigma_3$ vanish.
For $E_3 \to 0$ the  $nnd$ recombination rate \eqref{eq:K3} becomes
\begin{equation} \label{eq:K30}
K_3^0 = \frac{ 4\pi}{g_3}  (2\pi)^7 \mu_1 p_1 
 \sum_{m_s} | \langle \Phi_{3}^{0} |  T_{3 1} | \Phi_{1} \rangle |^2 ,
\end{equation}
where for the channel state $|\Phi_{3}^{0}  \rangle$
the relative momenta $k_y = k_z = 0$. The most
convenient representation for $|\Phi_{3}^{0}  \rangle$ is
 a single-component 2+2 state with $l_y=l_z=S_y=j_y=0$
and $j_x=S_z=\mathcal{J}=1$.

We study the four-nucleon system using  realistic high-precision
two-nucleon potentials, namely, the inside-nonlocal outside-Yukawa
(INOY04) potential  by Doleschall \cite{doleschall:04a,lazauskas:04a},
the Argonne (AV18) potential  \cite{wiringa:95a},
the charge-dependent Bonn potential (CD Bonn)  \cite{machleidt:01a},
and its extension CD Bonn + $\Delta$ \cite{deltuva:03c}
allowing for an excitation of a nucleon to a $\Delta$ isobar
and thereby yielding effective three- and four-nucleon forces.
Among these potentials only INOY04 nearly reproduces experimental binding energy of 
$\Hh$ (8.48 MeV), while AV18, CD Bonn and  CD Bonn + $\Delta$ 
underbind the $\Hh$  nucleus by 0.86, 0.48 and 0.20 MeV, respectively.

\begin{figure}[!]
\begin{center}
\includegraphics[scale=0.66]{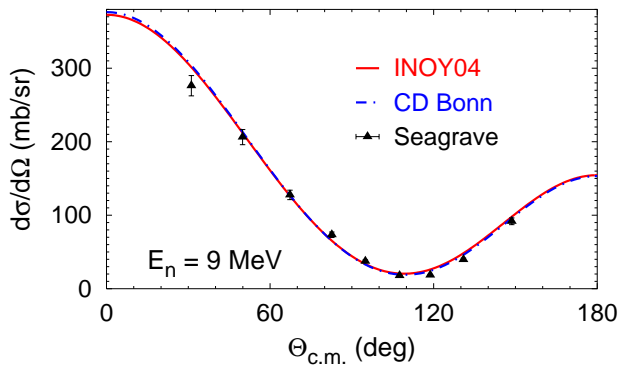}
\end{center} 
\caption{ \label{fig:dcs} (Color online) 
Differential cross section for elastic $n$-$\Hh$ scattering at 
9 MeV neutron energy as a function of c.m. scattering angle.
Results obtained with INOY04 (solid curves) and CD Bonn (dashed-dotted curves)
potentials are compared with the experimental data from
Ref.~\cite{seagrave:72}.}
\end{figure}

First we study the $\nH$ reactions for existing experimental data.
We concentrate on the energy regime relevant for the 
$nnd$ recombination, i.e., between the three- and four-cluster breakup
thresholds. In Fig.~\ref{fig:dcs} we show the differential
cross section for $n$-$\Hh$ elastic scattering at $E_n =9$ MeV neutron
energy corresponding to $E_3 = 0.49$ MeV.
The predictions agree well with the experimental data of
Ref.~\cite{seagrave:72} and 
are quite insensitive to the choice of the potential.
Results for $n$-$\Hh$ elastic scattering above 
the  four-cluster breakup threshold up to $E_n =22.1$ MeV
are given in Ref.~\cite{deltuva:12c}.

\begin{figure}[!]
\begin{center}
\includegraphics[scale=0.66]{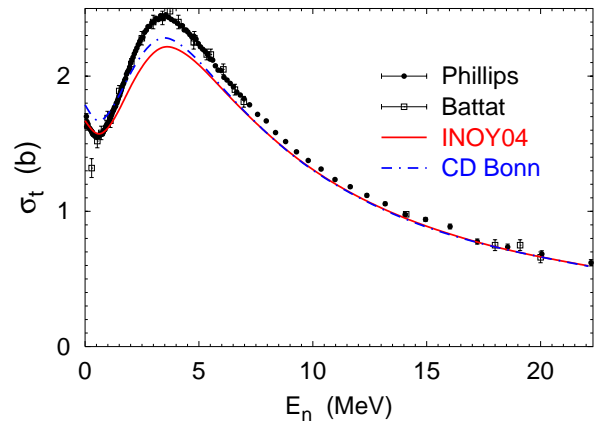}
\end{center} 
\caption{ \label{fig:tcs} (Color online) 
Total cross section for $n$-$\Hh$ scattering as a  function of
the neutron lab energy. 
Results obtained with INOY04 (solid curves) and CD Bonn (dashed-dotted curves)
potentials are compared with the experimental data from
Refs.~\cite{battat:59,phillips:80}.}
\end{figure}

In Fig.~\ref{fig:tcs} we show the total cross section for $n$-$\Hh$ 
scattering at neutron energies ranging from 0 to 22 MeV
and compare it to the data of Refs.~\cite{battat:59,phillips:80}.
The three-cluster (four-cluster) breakup threshold corresponds to
$E_n = 8.35$ (11.31) MeV. As already found in Refs.
\cite{lazauskas:04a,lazauskas:05a,deltuva:07a}, the total $n$-$\Hh$ 
cross section around the low-energy peak is underpredicted by the 
traditional two-nucleon potentials while the low-momentum
or chiral effective field theory potentials may provide a better
description \cite{deltuva:08b,viviani:fb19}.
Although with increasing energy the predictions approach the
experimental data, as already mentioned \cite{deltuva:07a},
the elastic and total cross section data may be inconsistent.
In the low-energy regime where the inelastic cross section should
 vanish for $E_n \le 8.35$ MeV and remain very small at $E_n=9$ MeV,
there is in general a better agreement
between predictions and experiment for the elastic differential
cross section than for the total cross section which
is significantly underestimated by theory.
A solution to this discrepancy may require new measurements
in this energy regime.

\begin{figure}[h]
\begin{center}
\includegraphics[scale=0.66]{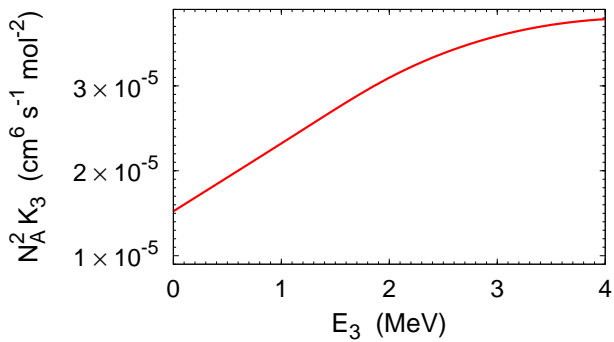}
\end{center} 
\caption{ \label{fig:k3} (Color online) 
$nnd$ recombination rate  $K_3$ as a function of 
relative kinetic $nnd$ energy $E_3$. 
Predictions are obtained using  the INOY04  potential.}
\end{figure}

In Fig.~\ref{fig:k3} we study the energy-dependence of  
the $nnd$ recombination rate in the standard form
$N_A^2 K_3$ where $N_A$ is the Avogadro's number.
We show only INOY04 predictions
as it is the only used potential with correct
$\epsilon_1$ and $p_1$ values. The results at $E_3=0$ 
are obtained from   Eq.~\eqref{eq:K30} while at $E_3 > 0$
it was more convenient to use  Eq.~\eqref{eq:K3s3} where
$\sigma_3$ was calculated using optical theorem.
Thus, for $E_3 > |\epsilon_d|$ our predictions in Fig.~\ref{fig:k3} 
estimate the upper limit for $N_A^2 K_3$ since they assume that 
the four-cluster breakup cross section is much smaller than the 
three-cluster breakup cross section. 
In the relevant energy regime  $0 \le E_3  \le |\epsilon_d|$
the $nnd$ recombination rate increases with increasing energy $E_3$
nearly linearly due to the increasing contributions of partial waves 
with nonzero orbital angular momentum $l_z$. 
The threshold values
$N_A^2 K_3^0$ referring to all employed potentials
are collected in  Table \ref{tab:k30}; they increase with
$\Hh$ binding energy.

\begin{table}[!]
\begin{ruledtabular}
\begin{tabular}{l*{2}{c}} & $|\epsilon_1|$ (MeV) 
&  $N_A^2 \,K_3^0 \; (\mathrm{cm}^6\mathrm{s}^{-1}\mathrm{mol}^{-2})$\\
\hline
AV18 &    7.62 & $1.31 \times 10^{-5}$ \\
CD Bonn & 8.00 & $1.41 \times 10^{-5}$ \\
CD Bonn + $\Delta$ & 8.28 & $1.47 \times 10^{-5}$ \\
INOY04 &  8.49 & $1.52 \times 10^{-5}$ \\
\end{tabular}
\end{ruledtabular}
\caption{ $nnd$ recombination rate at threshold 
calculated with different two-nucleon potentials.
The values for $\Hh$ binding energy are listed as well.}
\label{tab:k30}
\end{table} 

Finally we compare the relative importance of the $nnd$ recombination
and $nd$ radiative capture.
 For the latter the number of events,
i.e., the number of produced $\Hh$ nuclei per volume and time is
$K_2 \rho_n \rho_d$ with $K_2$ being the  $nd$ capture rate.
 The threshold value for it given in Ref.~\cite{fowler:67} is
$N_A K_2^0 = 66.2 \, \mathrm{cm}^3\mathrm{s}^{-1}\mathrm{mol}^{-1}$.
 The critical density of neutrons at which both processes yield
comparable contributions to the  $\Hh$ production  in the low-energy
(low-temperature) limit is given by  
$\rho_n^c = K_2^0/K_3^0 \approx 2.6 \times 10^{30} \, \mathrm{cm}^{-3}$.
This corresponds to the mass density of 
$4.4\times 10^{6} \, \mathrm{g}/\mathrm{cm}^{3}$. Thus, one may conclude
that at the neutron density available in the laboratories
(such as National Ignition Facility with expected
$\rho_n \sim 10^{22}$ to $10^{25} \, \mathrm{cm}^{-3}$
\cite{frenje:pc}) the
$nnd$ recombination is entirely irrelevant as well as for the
big-bang nucleosynthesis where the estimated baryon density
is even lower.
On the other hand, the neutron density in core-collapse supernova
or neutron stars  may be
higher than $\rho_n^c$ by several orders of magnitude
but the absence of deuterons renders 
$n+d$ and $n+n+d$ reactions irrelevant. 
However, based on our results one may conjecture that at such high densities 
the three-cluster recombination
of two neutrons and a heavier nucleus $A$, i.e.,
$n+n+A \to n + (An)$ might be as important as the
corresponding radiative capture $n+A \to \gamma + (An)$.
For example, the above reactions with $A$ being ${}^{20}$Ne
are relevant for the neon-burning process.
\\

\section{Summary \label{sec:sum}}

We have solved the four-nucleon AGS equations in the energy regime 
above the three-cluster
threshold and studied the rate for the recombination reaction
 $n+n+d \to n+\Hh$. The obtained results show that the 
$nnd$ recombination is not competitive with 
the radiative $nd$ capture for the production of tritium at 
 neutron densities available in laboratory induced fusion
or astrophysical processes. Thus, one may conjecture with a confidence
that other nucleon-nucleon-deuteron recombination reactions
(for example, $p+p+d \to p + \He$ that could contribute to the
hydrogen burning process in stars),
being in addition suppressed by the Coulomb repulsion, are inferior
to the respective nucleon-deuteron radiative capture reactions
at realistic densities, and that four-nucleon recombination reactions 
are even far less relevant.

In addition, we presented results for the $\nH$ elastic differential
cross section at $E_n = 9$ MeV and the $\nH$
total cross section up to $E_n = 22$ MeV. While the 
elastic cross section agrees fairly well with the data, there is a 
disagreement for the total cross section, especially in the low-energy regime.
This indicates a possible inconsistencies between the  $\nH$ 
elastic and  total cross section data and calls for
new measurements.

\begin{acknowledgments}
The authors thank J.~A. Frenje for discussions.
\end{acknowledgments}


\end{document}